\documentstyle [psfig,epsf] {l-aa}  
\def\infig#1#2#3{\epsfxsize=#3cm \centering{\mbox{\epsfbox{#2}}}}

\begin{document}

\thesaurus{06(19.34.1; 19.41.1; 07.16.1)}
 
\title{Grids of stellar models.}

\subtitle{VIII. From 0.4 to 1.0 ${\rm M_{\odot}}$ at Z=0.020 and Z=0.001, 
               with the MHD equation of state 
\thanks{Data available
at the CDS via anonymous ftp to
cdsarc.u-strasbg.fr (130.79.128.5) or via
http://cdsweb.u-strasbg.fr/Abstract.html } }

\author{C.Charbonnel 
        \inst{1,2}
        \and
        W.D\"appen
        \inst{3}
        \and
        D.Schaerer
        \inst{1,2}
        \and
        P.A.Bernasconi 
        \inst{4}
	\and 
        A.Maeder 
        \inst{4}
        \and
        G.Meynet 
        \inst{4}
        \and
        N.Mowlavi
        \inst{4}}

\offprints{C. Charbonnel}
\institute{Laboratoire d'Astrophysique de Toulouse, CNRS UMR 5572, 
14, Av. E. Belin, 31400 Toulouse, France
          \and
          Space Telescope Science Institute, 3700 San Martin Drive, 
          Baltimore, MD 21218, U.S.A.
          \and 
          Department of Physics and Astronomy, 
          University of Southern California, Los Angeles, CA 90089-1342, U.S.A.
          \and 
          Geneva Observatory, CH-1290 Sauverny, Switzerland}

\date{Received, accepted October 23, 1998}
\maketitle

\begin{abstract}

We present stellar evolutionary models covering the mass range from 0.4
to 1 M$_{\odot}$ calculated for metallicities Z=0.020 and 0.001 with the
MHD equation of state (Hummer \& Mihalas, 1988; Mihalas et al. 1988; 
D\"appen et al. 1988). A parallel calculation using the OPAL (Rogers et
al. 1996) equation of state has been made to demonstrate the adequacy of
the MHD equation of state in the range of 1.0 to 0.8 M$_{\odot}$ (the
lower end of the OPAL tables). Below, down to 0.4 M$_{\odot}$, we have
justified the use of the MHD equation of state by theoretical arguments
and the findings of Chabrier \& Baraffe (1997).

We use the radiative opacities by Iglesias \& Rogers (1996), 
completed with the atomic and molecular opacities by Alexander \&
Fergusson (1994). 
We follow the evolution from the Hayashi fully convective configuration 
up to the red giant tip for the most massive stars, and up to an age 
of 20 Gyr for the less massive ones.
We compare our solar-metallicity models with recent models computed by
other groups and with observations.

The present stellar models complete the set of grids computed with the same 
up-to-date input physics by the Geneva group 
[Z=0.020 and 0.001, Schaller et al. (1992), Bernasconi (1996), and 
Charbonnel et al. (1996); Z=0.008, Schaerer et al. (1992);
Z=0.004, Charbonnel et al. (1993); Z=0.040, Schaerer et al. (1993);
Z=0.10, Mowlavi et al. (1998);
enhanced mass loss rate evolutionary tracks, Meynet et al. (1994)]. 

\end{abstract}

\section{Introduction}

In stellar evolution computations, and in particular in the case of stars more
massive than the Sun, it is generally sufficient to use a simple equation of
state. The plasma of the stellar interior is treated as a mixture of perfect
gases of all species (atoms, ions, nuclei and electrons), and the Saha
equation is solved to yield the degrees of ionization or molecular formation.
In the case of low mass stars however, non ideal effects, such as Coulomb
interactions become important. It is then necessary to use a more adequate
equation of state than the one employed in the Geneva code for more
massive stars. This simple equation of state essentially contains a mixture 
of ideal gases, ionization of the chemicals is dealt with by the Saha equation, 
excited states and molecules are neglected, complete pressure ionization is 
artificially imposed above certain temperatures and pressures, and no 
Coulomb-pressure correction is included (see Schaller et al. 1992). 

For the present grids of models of 0.4 to 1.0 M$_{\odot}$ stars these
assumptions are obviously inadequate. For such stars, the most useful
equations of state, as far as their smooth realization and versatility are
concerned, are (i) the so-called Mihalas-Hummer-D\"appen (MHD) equation of
state (Hummer and Mihalas 1988; Mihalas et al. 1988; D\"appen et al. 1988),
and (ii) the OPAL equation of state, the major alternative approach developed
at Livermore (Rogers 1986, and references therein; Rogers et al. 1996). A
brief description of these two equations of state is given in the next
section.

Here, we chose the MHD equation of state. First, we were able to compute very
smooth tables specifically for our cases of chemical composition, instead of
relying on pre-computed, relatively coarse tables that would require
interpolation in the chemical composition. Second, our choice was forced by
the fact that the currently available OPAL equation of state tables do not
allow to go below stars less massive than $\sim$0.8 M$_{\odot}$. Third, we
validated our choice by a comparative calculation with OPAL at its low-mass
end. We found results that are virtually indistinguishable from MHD. Fourth,
we examined in a parallel theoretical study (Trampedach \& D\"appen 1998) the
arguments about the validity of the MHD equation of state down to the limit of
our calculation of 0.4 M$_{\odot}$ (see below).
Therefore we do not have to include a harder
excluded-volume term such as the one included in the
Saumon-Chabrier (SC) equation of state (Saumon \& Chabrier 1991, 1992).

Although the MHD equation of state was originally designed to provide the
level populations for opacity calculations of stellar {\it envelopes}, the
associated {\it thermodynamic quantities} of MHD can none the less be reliably
used also for stellar cores. This is due to the fact that in the deeper
interior the plasma becomes virtually fully ionized. Therefore, in practice,
it does not matter that the condition to apply the detailed Hummer-Mihalas
(1988) occupation formalism for bound species is not fulfilled, because
essentially there are no bound species. Other than that, the MHD equation of
state includes the usual Coulomb pressure and electron degeneracy, and can
therefore be used for low-mass stars and, in principle, even for envelopes
of white dwarfs (W. Stolzmann, {\it private communication}). The present
paper, with its MHD-OPAL comparison (see \S 3) corroborates this assertion.

This broad applicability of the MHD equation of state for entire stars was
specifically demonstrated by its successes in solar modeling and
helioseismology (Christensen-Dalsgaard et al. 1988, Charbonnel \& Lebreton
1993, Richard et al. 1996, Christensen-Dalsgaard et al. 1996). A solar model
that is based on the MHD equation of state from the surface to the center is
in all respects very similar to one based on the OPAL equation of state, the
major alternative approach developed at Livermore (Rogers 1986, and
references therein; Rogers et al. 1996). This similarity even pertains to the
theoretical oscillation frequencies that are used in comparisons with the
observed helioseismic data. 

Although the difference between the MHD and OPAL equations of state is of
helioseismological relevance, it has no importance for the lower-mass
stellar modeling of the present analysis. This is explicitly validated in 
the present paper. For the much finer helioseismological analyses,
it turned out that in some respect the OPAL model seems to be closer than 
the MHD model to the one inferred from helioseismological observations 
(Christensen-Dalsgaard et al. 1996, Basu \& Christensen-Dalsgaard 1997). 
However, we stress that for the present stellar modelling these subtle 
differences are no compelling reason to abandon the convenience of our 
ability to compute MHD equation of state tables ourselves, and to go below 
the range of the available OPAL tables ($\sim$0.8 M$_{\odot}$).

Not only helioseismology, but also fine features in the Hertzsprung-Russell
diagram of low- and very low-mass stars impose strong constraints on stellar
models (Lebreton \& D\"appen 1988, D'Antona \& Mazzitelli 1994, 1996, Baraffe
et al. 1995, Saumon et al. 1995). They have all confirmed the validity of the
principal equation-of-state ingredients employed in MHD (Coulomb pressure,
partial degeneracy of electrons, pressure ionization). Finally, we have
checked that even at the low-mass end of our calculations the physical
mechanism for pressure ionization in the MHD equation of state is still
achieved by the primary pressure ionization effect of MHD (the reduction of
bound-state occupation probabilities due to the electrical microfield; see
Hummer \& Mihalas 1988). Such a verification was necessary to be sure that our
results are not contaminated by the secondary, artificial pressure-ionization
device included in MHD for the very low-temperature high-density regime (the
so-called $\Psi$ term of Mihalas et al. 1988). A parallel calculation has
confirmed that in our models a contamination by this $\Psi$ term can be ruled
out (Trampedach \& D\"appen 1998).

In the present paper, we expand the current mass range of the Geneva evolution
models from 0.8 down to 0.4 M$_{\odot}$, by using a specifically calculated
set of tables of the MHD equation of state. This work aims to complete the
base of extensive grids of stellar models computed by the Geneva
group with up-to-date input physics [Z=0.020 and 0.001, Schaller et al.
(1992), Bernasconi (1996), and Charbonnel et al. (1996); Z=0.008, Schaerer et
al. (1992); Z=0.004, Charbonnel et al. (1993); Z=0.040, Schaerer et al.
(1993); Z=0.10, Mowlavi et al. (1998); enhanced mass loss rate evolutionary
tracks, Meynet et al. (1994) ]. In Sect. 2, we present the characteristics of
our equation of state and recall the physical ingredients used in our
computations. In Sect. 3, we summarize the main characteristics of the present
models and discuss the influence of the equation of state on the properties of
low mass stars. Finally, we compare our solar-metallicity models with recent
models computed by other groups and with observations in \S 4.
\section{Input physics}

The basic physical ingredients used for the complete set of grids of the
Geneva group are extensively described in previous papers 
(see Schaller et al. 1992, hereafter Paper I). 
With the exception of the equation of state, described in detail
in the following subsection, we therefore just mention the main points.  

\subsection{Equation of state}

We have already justified our choice of the MHD equation of state in the
introduction. As mentioned there, MHD is one of the two recent equations of
state that have been especially successful in modeling the Sun under the
strong constraint of helioseismological data. Historically, the MHD equation
of state was developed as part of the international ``Opacity Project'' [OP,
see Seaton (1987, 1992)]. It was realized in the so-called {\it chemical
picture}, where plasma interactions are treated with modifications of atomic
states, {\it i.e.} the quantum mechanical problem is solved before statistical
mechanics is applied. It is based on the so-called free-energy minimization
method. This method uses approximate statistical mechanical models (for
example the nonrelativistic electron gas, Debye-H\"uckel theory for ionic
species, hard-core atoms to simulate pressure ionization via configurational
terms, quantum mechanical models of atoms in perturbed fields, etc). From
these models a macroscopic free energy is constructed as a function of
temperature $T$, volume $V$, and the concentrations $N_1, \ldots, N_m$ of the
$m$ components of the plasma. The free energy is minimized subject to the
stoichiometric constraint. The solution of this minimum problem then gives
both the equilibrium concentrations and, if inserted in the free energy and
its derivatives, the equation of state and the thermodynamic quantities.

The other of these two equations of state is the one underlying the OPAL
opacity project (see \S 2.2). The OPAL equation of state is
realized in the so-called {\it physical picture}. It starts out from the grand
canonical ensemble of a system of the basic constituents (electrons and
nuclei), interacting through the Coulomb potential. Configurations
corresponding to bound combinations of electrons and nuclei, such as ions,
atoms, and molecules, arise in this ensemble naturally as terms in cluster
expansions. Any effects of the plasma environment on the internal states are
obtained directly from the statistical-mechanical analysis, rather than by
assertion as in the chemical picture. 

More specifically, in the chemical picture, perturbed atoms must be introduced
on a more-or-less {\it ad-hoc} basis to avoid the familiar divergence of
internal partition functions (see {\it e.g.} Ebeling et al. 1976).
In other words, the approximation of unperturbed atoms precludes the
application of standard statistical mechanics, {\it i.e.} the attribution of a
Boltzmann-factor to each atomic state. The conventional remedy of the chemical
picture against this is a modification of the atomic states, {\it e.g.} by
cutting off the highly excited states in function of density and temperature
of the plasma. Such cut-offs, however, have in general dire consequences due
to the discrete nature of the atomic spectrum, {\it i.e.} jumps in the number
of excited states (and thus in the partition functions and in the free energy)
despite smoothly varying external parameters (temperature and density).
However, the occupation probability formalism employed by the MHD equation of
state avoids these jumps and delivers very smooth thermodynamic quantities.
Specifically, the essence of the MHD equation of state is the Hummer-Mihalas
(1988) occupation probability formalism, which describes the reduced
availability of bound systems immersed in a plasma. Perturbations by charged
and neutral particles are taken into account. The neutral contribution is
evaluated in a first-order approximation, which is good for stars in which
most of the ionization in the interior is achieved by temperature [the
aforementioned study (Trampedach \& D\"appen 1998) has verified the
validity of this assumption down to the lowest mass of our calculation].
For colder objects (brown dwarfs, giant planets), higher-order excluded-volume
effects become very important (Saumon \& Chabrier 1991, 1992; Saumon et
al. 1995). In the common domain of application of the Saumon et al.
(1995) and MHD equations of state, Chabrier \& Baraffe (1997) showed 
that both developments yield very similar results, which strongly
validates the use of the MHD equation of state for our mass range of 0.4
to 1.0 M$_{\odot}$.

Despite undeniable advantages of the physical picture, the chemical picture
approach leads to smoother thermodynamic quantities, because they can be
written as analytical (albeit complicated) expressions of temperature, density
and particle abundances. In contrast, the physical picture is normally
realized with the unwieldy chemical potential as independent variable, from
which density and number abundance follow as dependent quantities. The
physical-picture approach involves therefore a numerical inversion before the
thermodynamic quantities can be expressed in their ``natural'' variables
temperature, density and particle numbers. This increases computing time
greatly, and that is the reason why so far only a limited number of OPAL
tables have been produced, only suitable for stars more massive than $\sim$0.8
M$_{\odot}$. Therefore we chose MHD for its smoothness, availability, and the
possibility to customize it directly for our calculation, despite the -- in
principle -- sounder conceptual foundation of OPAL.

\subsection{Opacity tables, treatment of convection, atmosphere
and mass loss}

\begin{itemize}
\item The OPAL radiative opacities from Iglesias \& Rogers (1996) 
including the spin-orbit interactions in Fe and relative metal abundances
based on Grevesse \& Noels (1993) are used. 
These tables are completed at low temperatures below 10000 K with the 
atomic and molecular opacities by Alexander \& Fergusson (1994).

\item We use a value of 1.6 for the mixing length parameter $\alpha$.  
Various observational comparison support this choice. 
$\alpha = 1.6 \pm 0.1$ leads to the best fit of the red giant branch for 
a wide range of clusters (see Paper I). It is also the value we
get for the calibration of solar models including the same input
physics (Richard et al. 1996). 

\item A grey atmosphere in the Eddington approximation is adopted as
boundary condition. Below $\tau = 2/3$, full integration of the
structure equations is performed. 
We discuss the implications of such an approximation in \S 4.

\item Evolution on the pre-main sequence and on the main sequence are 
calculated at constant mass. 
On the red giant branch, we take mass loss into account by using the 
expression by Reimers (1975) : 
$\dot{M} = 4 \times 10^{-13} \eta L R / M$ (in M$_{\odot}$yr$^{-1}$) where
L, M and R are the stellar luminosity, mass and radius respectively (in 
solar units). At solar metallicity, $\eta=$0.5 is chosen (see Maeder \& 
Meynet 1989). At Z=0.001, the mass loss is lowered by a factor 
(0.001/0.020)$^{0.5}$ with respect to the models at Z=0.020 for the same 
stellar parameters. 

\subsection{Nuclear reactions}

\item Nuclear reaction rates are due to Caughlan \& Fowler (1988).
The screening factors are included according to the prescription of 
Graboske et al. (1973).

\item Deuterium is destroyed on the pre-main sequence 
at temperatures higher than 10$^6$ K by 
D(p,$\gamma)^3$He and, to a lower extent, 
by D(D,p)$^3$H(e$^- \nu)^3$He and D(D,n)$^3$He. 
We take into account these three reactions. 
In order to avoid the follow-up of tritium, 
we consider the last two reactions as a single process, D(D,nucl)$^3$He. 
The $\beta$ desintegration is considered as instantaneous, which is 
justified in view of the lifetime of tritium ($\tau _{1/2}$=12.26 yr) 
compared with the evolutionary timescale.
The rate of the D(D,nucl)$^3$He reaction is written as 
$$<DD>_{nucl} = (1 + {{<DD>_p}\over{<DD>_n}} ) <DD>_n $$
\noindent
where we take for ${{<DD>_p}\over{<DD>_n}}$ a mean value of 1.065, 
in agreement with the rates given by Caughlan \& Fowler (1988).
The corresponding mean branch ratios are 
I$_p$ = 0.5157 and I$_n$ = 0.4843.
\end{itemize}

\subsection{Initial abundances}

\begin{itemize}

\item The initial helium content is determined by 
Y=0.24+($\Delta$Y/$\Delta$Z)Z, 
where 0.24 corresponds to the current value of the cosmological 
helium (Audouze 1987).
We use the value of 3 for the average relative ratio of helium to metal
enrichment ($\Delta$Y/$\Delta$Z) during galactic evolution.
This leads to (Y,Z) = (0.300,0.020) and (0.243,0.001).
In addition, computations were also performed with (Y,Z) = (0.280,0.020).

\item The relative ratios for the heavy elements correspond to the 
mixture by Grevesse \& Noels (1993) used in the opacity computations by 
Iglesias \& Rogers (1996).

\item Choosing initial abundance values for D and $^3$He is more complex.
Pre-solar abundances for both elements have been reviewed in Geiss (1993), 
however galactic chemical models face serious problems to describe their  
evolution (see e.g. Tosi 1996), and no reliable prescription 
exists to extrapolate their values in time. 
Deuterium is only destroyed by stellar processing since the Big Bang
Nucleosynthesis, so that its abundance decreases with time (i.e. with increasing metallicity). 
On the other hand, 
the actual contribution of stars of different masses is still subject 
to a large debate (Hogan 1995, Charbonnel 1995, Charbonnel \& Dias 1998), 
and observations of ($^3$He/H) in the proto-solar nebulae (Geiss 1993) and in 
galactic HII regions present a large dispersion. 
We adopt the same (D/H) and ($^3$He/H) initial values for both metallicities, 
namely 2.4$\times 10^{-5}$ and 2.0 $\times 10^{-5}$ respectively.
\end{itemize}

\subsection{Initial models}

At the low mass range considered in this paper, the observed quasi-static
contraction begins rather close to the theoretical deuterium main sequence,
once the stars emerge from their parental dense gas and dust. For the
present purposes then, we take as starting models polytropic
configurations
on the Hayashi boundary, neglecting the corrections brought to
isochrones and upper tracks by the modern accretion paradigm of star
formation (Palla \& Stahler 1993, Bernasconi \& Maeder 1996). 
For the mass range considered here, these corrections are likely not to
exceed 3$\%$ of the Kelvin-Helmholtz timescale for the pre-main sequence
contraction times (Bernasconi 1996). 
We note, however, that the predicted upper locus for the optical appearance 
of T~Tauri stars in the HR diagram can be as much as half less luminous than 
the deuterium ignition luminosity on the convective tracks
($\Delta \log L \approx$ 0.3).
\def\tiret{\multicolumn{2}{c}{\mbox{---}}}

\section{Short discussion of the main results}

\subsection{HR diagram and lifetimes}

\begin{figure}
\infig{8cm}{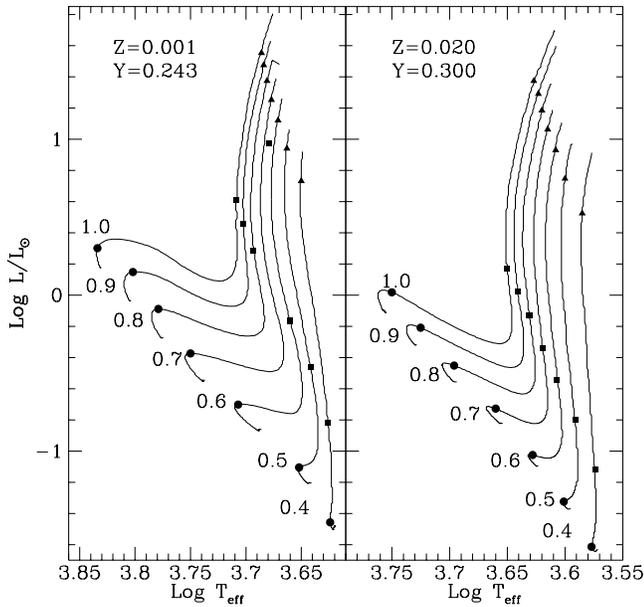}{8.8}
\caption{Theoretical HR diagrams for pre-main sequence evolution for all 
our models (from 0.4 to 1.0 M$_{\odot}$) for both metallicities. 
Triangles mark the ignition of deuterium
burning. The radiative core appears at the square location, and the
development of a small convective core is indicated by the circles (for
Z=0.020, the tracks are given only for the models with Y=0.30 to avoid
confusion). 
}
\end{figure}

\begin{figure}
\infig{8cm}{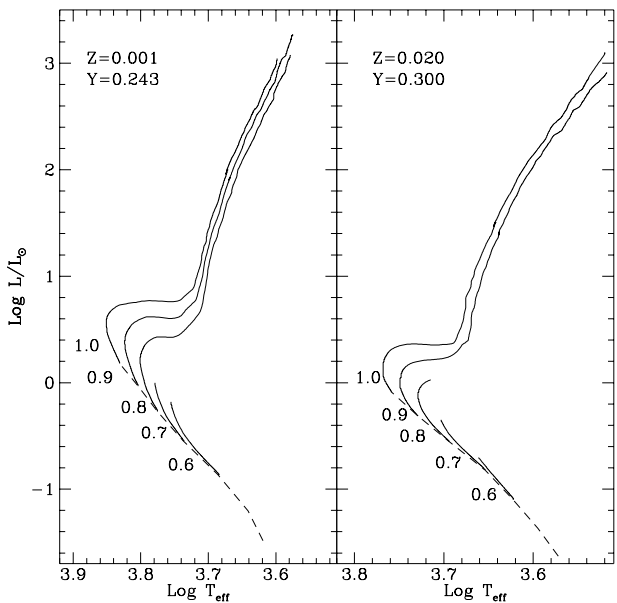}{8.8}
\caption{Theoretical HR diagrams for stellar masses between 0.6 and
1M$_{\odot}$, and position of the zero age main sequence (dashed line)
for the entire mass interval.}
\end{figure}

The HR diagrams for pre-main sequence evolution and for the following phases
are given in Fig. 1 and 2 respectively for both metallicities. For each
stellar mass, Table 1 displays the lifetimes in the contraction phase and in
the deuterium- and hydrogen-burning phases. Note that we did not complete the
main sequence evolution computations for the less massive stars which have a
H-burning phase longer than the age of the universe; for these stars, our last
computed model corresponds to an age of 20 Gyr. 

\begin{itemize}
\item In the stellar mass range we consider, the ignition of deuterium 
burning
(indicated in Fig.1) takes place in a fully convective interior. 
During pre-main sequence evolution, a radiative core develops. 
However a proper radiative branch is absent, since these stars maintain a 
convective envelope all along contraction until the ZAMS has been reached, 
and further on. 
\item From Table 1 one can see that the contraction time lasts less than 
4 thousandths of the hydrogen-burning stage.
\item Due to opacity effects, the entire evolution (pre-main sequence, 
main sequence and red giant branch) occurs at higher luminosity and 
effective temperature for a given stellar mass when the initial metallicity 
is smaller, or when the hydrogen content is lower for the same
initial metallicity. 
\item As a consequence, the contraction phase and the deuterium- and 
hydrogen-burning phases for a given stellar mass are shorter at lower 
metallicity (see Table 1), and at lower hydrogen content for the same
value of Z. 
\end{itemize}

\subsection{Influence of the equation of state}

\begin{figure}
\infig{8cm}{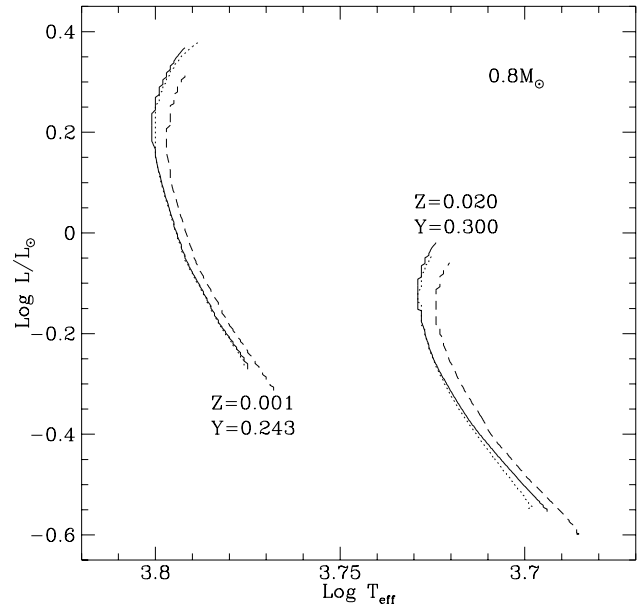}{8.8}
\caption{Influence of the MHD equation of state on the main sequence 
evolutionary track of the 0.8M$_{\odot}$ models. The solid, dotted and dashed 
lines correspond to models computed with the MHD, OPAL and the
simple Geneva (section~1) equations of state, respectively.}
\end{figure}

\begin{figure}
\infig{8cm}{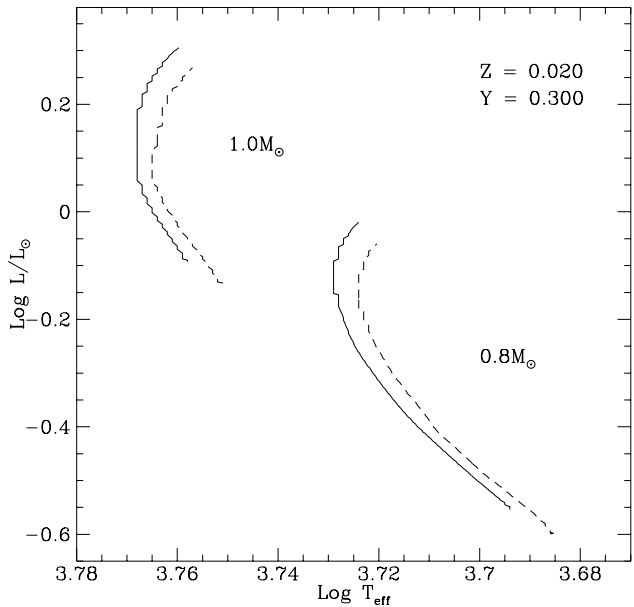}{8.8}
\caption{Influence of the MHD equation of state on the main sequence 
evolutionary track of the 1.0 and 0.8M$_{\odot}$ models for Z=0.020. 
The full and dashed lines correspond to models computed with the MHD and with 
the simple Geneva equations of state, respectively.
}
\end{figure}

\begin{table}
\caption{Lifetimes in contraction and nuclear phases
(in units of $10^6$ yr), and ratio of the contraction time t$_c$ to the
hydrogen burning time t$_H$. For the less massive stars which have a main
sequence phase longer than the Hubble time, we stopped the computations at an
age of 20 Gyr.}
$${\begin{array}{r@{.}lr@{.}lr@{.}lr@{.}lr@{.}lr@{.}l}  \hline \\[1mm]
\multicolumn{2}{c}{\mbox{Z}}&
\multicolumn{2}{c}{\mbox{Initial}}&
\multicolumn{2}{c}{\mbox{Contraction}} &
\multicolumn{2}{c}{\mbox{D-burning}} &
\multicolumn{2}{c}{\mbox{H-burning}}&
\multicolumn{2}{c}{\mbox{ {t$_c$ / t$_H$} }} \\
\multicolumn{2}{c}{\mbox{Y}}&
\multicolumn{2}{c}{\mbox{mass}}   &
\multicolumn{2}{c}{\mbox{phase}} &
\multicolumn{2}{c}{\mbox{phase}} &
\multicolumn{2}{c}{\mbox{phase}} &
\multicolumn{2}{c}{} \\[1mm]
 \hline
\\[0.5mm]
0&001 & 0&4 & 160&10  & 0&417 & \tiret &  \tiret  \\
0&243 & 0&5 &  81&11  & 0&321 & \tiret &  \tiret  \\
 &    & 0&6 &  51&89  & 0&277 & \tiret &  \tiret  \\
 &    & 0&7 &  36&44  & 0&240 & \tiret &  \tiret  \\
 &    & 0&8 &  31&20  & 0&214 & 14338&03 & 0&0022 \\
 &    & 0&9 &  22&82  & 0&196 &  9051&01 & 0&0025 \\
 &    & 1&0 &  19&34  & 0&187 &  6847&20 & 0&0028 \\[1mm]
\hline
0&020 & 0&4 & 153&84  & 0&605 & \tiret &  \tiret  \\
0&300 & 0&5 & 117&34  & 0&476 & \tiret &  \tiret  \\
 &    & 0&6 &  88&26  & 0&389 & \tiret &  \tiret  \\
 &    & 0&7 &  68&80  & 0&330 & \tiret &  \tiret  \\
 &    & 0&8 &  58&15  & 0&302 & 22713&11 & 0&0026 \\
 &    & 0&9 &  44&30  & 0&276 & 14070&96 & 0&0031 \\
 &    & 1&0 &  32&80  & 0&252 &  9059&52 & 0&0036 \\[1mm]
\hline
0&020 & 0&4 & 161&79  & 0&640 & \tiret &   \tiret   \\
0&280 & 0&5 & 124&50  & 0&484 & \tiret &   \tiret   \\
 &    & 0&6 &  93&75  & 0&407 & \tiret &   \tiret   \\
 &    & 0&7 &  72&86  & 0&344 & \tiret &   \tiret   \\
 &    & 0&8 &  62&34  & 0&309 & 26372&72 & 0&0024 \\
 &    & 0&9 &  48&26  & 0&286 & 16421&94 & 0&00002 \\
 &    & 1&0 &  37&98  & 0&263 & 10662&62 & 0&00002 \\[1mm]
\hline
\end{array}} $$
\end{table}

\begin{table}
\caption{Lifetimes on the main sequence (in units of $10^6$ yr)
for models computed with the MHD, the OPAL and the simple Geneva equation of
state.}
\begin{tabular}{ccccc}
\hline \\[0.4mm]
\multicolumn{1}{c}{Z}&
\multicolumn{1}{c}{M/M$_{\odot}$} &
\multicolumn{1}{c}{t$_H$(MHD)}&
\multicolumn{1}{c}{t$_H$(OPAL)}&
\multicolumn{1}{c}{t$_H$(orig. Geneva)} \\[1mm]
\hline \\[1mm]
0.020 & 1.0 & 9059.5 & 8968.1  &  9970.1 \\
0.020 & 0.8 &22713.3 & 22361.4 & 25151.3 \\
0.001 & 0.8 &14338.0 & 13986.0 &         \\[1mm]
 \hline
\end{tabular}
\end{table}

When we first compare the results obtained with the MHD and with the 
simple Geneva equations of state (see Fig. 3 and 4, and Table 2), we 
obtain essentially the same results than Lebreton \& D\"appen (1988). 
Firstly, the
fact that MHD contains ${\rm H}_2$ molecules, and the simple Geneva code
does not, is reflected in a shift essentially {\it along} the ZAMS. On the
other hand, the Coulomb pressure correction, also contained in MHD, causes a
slight shift of the ZAMS, clearly visible for higher masses, where there are
no hydrogen molecules in the photosphere. This Coulomb effect has been well
discussed in the case of helioseismology ({\it e.g.} Christensen-Dalsgaard et
al. 1996). Conformal to the effect of the MHD equation of state to push the
apparent position on the ZAMS upward, it also decreases the lifetime on the
ZAMS (see Table 2).

For comparison, we have computed with the OPAL equation of state two
0.8M$_{\odot}$ models (the lowest mass that can be computed with the current
OPAL tables), for both metallicities. As can be seen in Fig.3, the
corresponding tracks are very close to those obtained with the MHD
equation of state, the use of the OPAL equation of state
leading to slightly higher effective temperature on the ZAMS. As far as
their internal structure is concerned, the models computed with MHD equation
of state have slightly deeper convection zones. The main sequence lifetime
obtained with the MHD equation of state is slightly higher than the one
obtained with the OPAL equation of state (Table 2). The comparison shows that
down to 0.8M$_{\odot}$ all is fine with the MHD pressure ionization. As
mentioned in the introduction, Trampedach \& D\"appen (1998) predict a correct
functioning of pressure ionization in MHD even for much smaller masses. With
the present comparison, we have validated their prediction at least to
0.8M$_{\odot}$.

\section{Comparisons with other sets of models and with observations}

We now compare our models with recent models computed by other groups and
with various observations. 

The strongest approximation in our models lies in the treatment of the 
atmosphere and of the surface boundary conditions,
which are specified by the Eddington approximation.
In Figs.\ 5 and 6 we compare our results with the models of D'Antona \&
Mazzitelli (1994) and Tout et al. (1996) which also rely on a simple 
treatment of the boundary conditions. For the mass range considered in our 
work we obtain predictions very similar to  D'Antona \& Mazzitelli.

Sophisticated model atmospheres for the computation of very low
mass stars have been developed recently (Baraffe et al. 1995, 1998; 
Brett 1995, Allard et al. 1997).
Their use becomes crucial for very low mass stars ($M \la 0.4$ M$_{\odot}$)
down to the brown dwarf limit. 
We refer to Chabrier \& Baraffe (1997) and Baraffe et al. (1998) for a 
discussion of the physical basis of the differences. 
As can be seen in Fig.5, the Eddington approximation we use results in
higher T${\rm eff}$ (from 2 to 5 $\%$ depending on the metallicity) 
compared to the Chabrier \& Baraffe (1997) models at our low mass end;
our models also have slightly smaller radius (see Fig.6).
In this mass range the predictions agree well with the observations
of Popper (1980) and Leggett et al. (1996) and do not allow 
to disregard one model with respect to the other.

\begin{figure}
\infig{8cm}{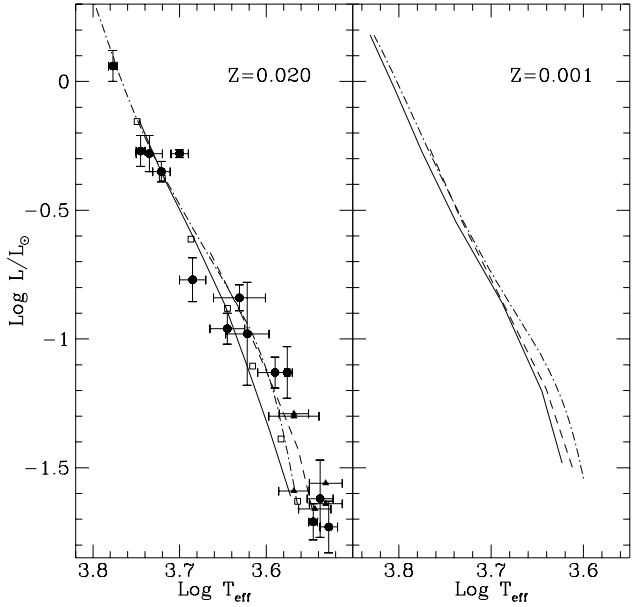}{8.8}
\caption{Theoretical HR diagrams at two values of the metallicity.
Solid lines : our models (Y=0.28 for Z=0.02); dashed lines : Chabrier \& 
Baraffe (1997); dashed-dotted lines : Tout et al. (1997); open squares : 
D'Antona \& Mazzitelli (1994). The models are taken at 0.1 Gyr, except for
the zams models of T97. 
The observational data are from Popper (1980; black points) and from
Leggett et al.(1996; black triangles).}
\end{figure}

\begin{figure}
\infig{8cm}{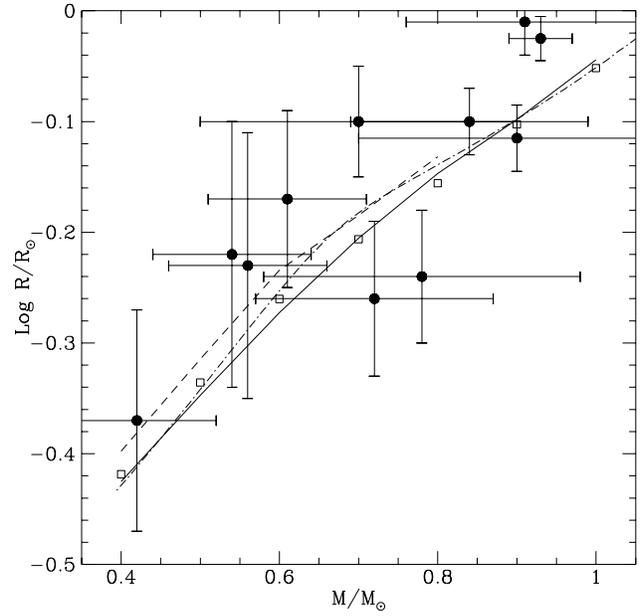}{8.8}
\caption{Dependence of the stellar radius on mass for solar metallicity
models. See Fig.5 for the references of the models and observations.}
\end{figure}

\begin{figure}
\infig{8cm}{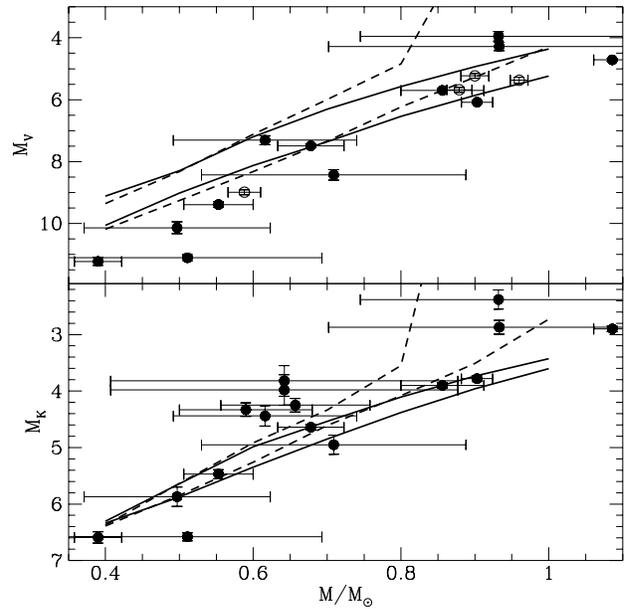}{8.8}
\caption{Mass-luminosity relations at 0.1 and 10 Gyr (solid and dashed
lines respectively) for the solar (with Y=0.28) and low metallicity models 
(lower and upper curves respectively).
The observations are from Andersen (1991) and Henry \& McCarthy (1993)
(white and black circles respectively)}
\end{figure}

In Fig.7 we show the predicted mass-luminosity relations for the V and K band
for different metallicities and ages. The magnitudes were derived from
the standard stellar library for evolutionary synthesis of Lejeune et al.\
(1998), which provides empirically calibrated colours for solar metallicity
and a semi-empirical correction for non-solar metallicities.
The comparison with the observations of Andersen (1991) and Henry \& McCarthy 
(1993) shows a good agreement with our predictions for both the V and K 
band (Fig.7). Again a similar agreement is obtained with the
models of Brocato et al.\ (1998) and Baraffe et al.\ (1998).

From the comparisons in Figs.5 to 8, we conclude (in agreement with
Alexander et al.\ 1997) that for the mass range
considered in this work an approximate treatment of the stellar atmosphere 
leads to a satisphactory agreement between theoretical predictions and 
observations. Independently ab initio stellar interior and atmosphere models 
allowing the detailed predictions of all observational properties are
of fundamental importance for our understanding of very low mass stars
which are out of the scope of the present grids.

\appendix

\section{Appendix: How to obtain the tables by file-transfer}
The results of this work and of our previous grids 
(Papers I to VI) are published by Astronomy and Astrophysics at the Centre 
de Donn\'ees Spatiales (CDS at Strasbourg) where the corresponding tables 
are available in electronic form: 

{\tt http://cdsweb.u-strasbg.fr}.
These data can also be obtained from the Geneva Observatory 

{\tt http://obswww.unige.ch/}
(contact ~Corinne.Charbonnel@obs-mip.fr).  

An ensemble of models were selected to describe each evolutionary track. 
For each model the tables display the age, actual mass,  log $L/L_\odot$, 
log $T_{eff}$, the surface abundances in mass fraction of H, $^4$He, $^{12}$C, 
$^{13}$C, $^{14}$N, $^{16}$O, $^{17}$O, $^{18}$O,  $^{20}$Ne, $^{22}$Ne,  
the core mass fraction Qcc, log$(-\dot{M})$ (where $\dot{M}$ is the mass loss 
rate on the red giant branch), log $\rho _c$ (where $\rho _c$ is the central 
density), log T$_c$ (where T$_c$ is the central temperature), and the central 
abundances in mass fraction of the above elements. A detailed description of 
the models selection and of the tables contents is given in Paper I. 
We now also provide photometric data in verious systems for all tracks 
and isochrones (see Schaerer \& Lejeune 1998).

\acknowledgements{We thank Thibault Lejeune for calculations prior
to publication.
C.C. wishes to express her gratitude to the staff
of the Space Telescope Science Institute where this study was carried out.
W.D. was supported in part by the grants AST-9315112 and AST-9618549 of
the National Science Foundation.
D.S. is supported by the Swiss National Foundation for Scientific Research
and acknowledges partial support from the Directors Discretionary Research 
Fund of the STScI.}

\end{document}